\pdfoutput=1

\documentclass[12pt]{iopart}
\usepackage{iopams} 
\usepackage{array,dsfont,graphicx,multirow,rotating,tensor,verbatim,youngtab}
\usepackage[bookmarksnumbered,bookmarksopen=true,bookmarksopenlevel=1,linktocpage,pdfstartview=FitH]{hyperref}
\usepackage[all]{hypcap}

\newcolumntype{M}[1]{>{$}{#1}<{$}}

\newcolumntype{M}[1]{>{$}{#1}<{$}}

\newcommand{\ket}[1]{|  #1 \rangle}
\newcommand{\bra}[1]{\langle  #1 |}

\newcommand{\sst}[1]{{\scriptscriptstyle #1}}

\newcommand{\rep}[1]{\mathbf{#1}}

\def\0{{\sst{(0)}}}
\def\1{{\sst{(1)}}}
\def\2{{\sst{(2)}}}
\def\3{{\sst{(3)}}}
\def\4{{\sst{(4)}}}
\def\5{{\sst{(5)}}}
\def\6{{\sst{(6)}}}
\def\7{{\sst{(7)}}}

\newtheorem{theorem}{Theorem}
\newtheorem{definition}{Definition}
\newtheorem{lemma}[theorem]{Lemma}

\newcommand{\Aut}{\textrm{Aut}}

\newcommand{\Det}{\textrm{Det}}

\newcommand{\Hom}{\textrm{Hom}}

\newcommand{\SL}{\textrm{SL}}

\newcommand{\Sym}{\textrm{Sym}}

\newcommand{\be}{\begin{equation}}
\newcommand{\ee}{\end{equation}}
\newcommand{\bes}{\begin{array}{lll}}
\newcommand{\bea}{\begin{eqnarray}}
\newcommand{\eea}{\end{eqnarray}}

\newcommand{\braket}[2]{\langle#1|#2\rangle}

\newcommand{\J}{\mathfrak{J}}

\newcommand{\F}{\mathds{F}}

\newcommand{\C}{\mathds{C}}

\newcommand{\FTS}{\mathfrak{F}}

\begin{document}

\title{Freudenthal  ranks: GHZ vs. W}

\author{L. Borsten}
 \address{Blackett Laboratory, Imperial College, London, SW7 2AZ, U.K.}
\ead{leron.borsten@imperial.ac.uk}
\begin{abstract}
The Hilbert space of three-qubit pure states may be identified with a Freudenthal triple system. Every state has an unique  \emph{Freudenthal rank} ranging from 1 to 4, which is determined by a set of  automorphism group covariants. It is shown here that the optimal success rates for winning a three-player non-local game, varying over all local strategies, are strictly ordered by the Freudenthal rank of the shared three-qubit resource.
\end{abstract}


\maketitle

\section{Introduction}

It is by now well known that, under the paradigm of stochastic local operations and classical communication (SLOCC), three qubits\footnote{Here and throughout we restrict our attention to pure states. An FTS perspective on three-qubit mixed state entanglement can be found in \cite{Szalay:2012}.}  can be entangled in four physically distinct ways: 1) Separable $A$-$B$-$C$, 2) Biseparable $A$-$BC$, 3) Totally entangled $W$ states, 4) Totally entangled GHZ states (Greenberger-Horne-Zeilinger) \cite{Dur:2000}. The most important and interesting aspect of this classification is the appearance of two inequivalent forms of totally entangled states, $W$ and GHZ. It is not enough to simply declare a state is totally entangled, one must also specify \emph{how} it is totally entangled. 

This three-qubit SLOCC classification may be  elegantly captured by identifying the three-qubit state space with a particular Freudenthal triple system (FTS) defined over a  cubic Jordan algebra \cite{Borsten:2009yb}. In the present work, this construction is reformulated  in \autoref{sec:fts} using the axiomatic  FTS, which dispenses with the underlying Jordan algebra. This FTS framework is not limited to  three qubits\footnote{Remarkably, the seemingly unrelated concept of Freudenthal duality, introduced in the context of supergravity \cite{Borsten:2009zy, Ferrara:2011gv,Galli:2012ji,Borsten:2012pd}, also has a qubit significance  \cite{Levay:2012tg}.}  and has been extended to a number of more exotic multipartite systems including mixtures of bosonic and fermionic qudits \cite{Borsten:2008, Borsten:2008wd, levay-2008, Levay:2009, Borsten:2012fx}.

An important  feature common to all FTS  is the universal  notion of   \emph{rank}. Any element of a given FTS  has a unique rank $1, 2, 3$ or $4$. For the  three-qubit FTS these ranks are nothing but the SLOCC entanglement classes: Rank 1) Separable $A$-$B$-$C$, Rank 2) Biseparable $A$-$BC$, Rank 3) Totally entangled $W$ states, Rank 4) Totally entangled GHZ states.

The labelling of the ranks 1 through 4 is not incidental; they are so ordered by implication through the defining rank conditions \eref{eq:FTSrank}. This suggests, from the perspective of the FTS, that $W$ and GHZ are not merely inequivalent, but in fact    ordered; GHZ is both differently and \emph{more} entangled than $W$, in some precise sense. 

This particular mathematical ordering of the entanglement classes naturally raises the question of its physical significance. What  set-up would lead three experimenters, with no knowledge of SLOCC, to conclude unequivocally that a black-box secretly containing a rank 4 state is more non-local than one containing a rank 3 state? Is there a single experiment which separates out all  the FTS ranks? 

It turns out that the obvious guess, Mermin's elegant three-party GHZ experiment \cite{Mermin:1990},  is also the correct guess. To make the logic as clear as possible  we adopt a reformulation of  Mermin's set-up in terms of a \emph{non-local cooperative game of incomplete information}  introduced in \cite{Watrous:2006}. In this language the contradiction with local realism exposed by Mermin   translates into the existence of a local strategy utilising the GHZ state that wins the game with certainty. 

Specifically, it is shown here that the algebraic Freudenthal rank conditions \emph{alone}  imply that there is \emph{no} local strategy utilising a rank 3  state that wins the game with certainty, in contrast to the rank 4 GHZ case. In fact, the optimal success rates are strictly ordered according as the rank: 
\be
3/4=p(\textrm{rank } 1)<p(\textrm{rank } 2)<p(\textrm{rank } 3)<p(\textrm{rank } 4)=1,
\ee
where $p(\textrm{rank } n)$ denotes the greatest possible probability of winning using a rank $n$ state.
On this basis we argue that the physical significance of the three-qubit Freudenthal ranks is most naturally expressed in terms of this three-player non-local game.

\section{Freudenthal SLOCC  classification}\label{sec:fts}

\subsection{The Freudenthal triple system}\label{sec:ftsdef}

In 1954 Freudenthal \cite{Freudenthal:1954, Freudenthal:1959} found that the
133-dimensional exceptional Lie group $E_{7}$ could be understood in terms
of the automorphisms of a construction based on the minuscule  56-dimensional $%
E_{7}$-module  built from the exceptional Jordan algebra of $%
3\times 3$ Hermitian octonionic matrices. Today this construction goes by
the name of \emph{the Freudenthal triple system}, reflecting the special
role played by its triple product.

Following Freudenthal, Meyberg \cite{Meyberg:1968} and Brown \cite%
{Brown:1969} axiomatized the ternary structure underlying the FTS. The $E_7$%
-module is just one of a class of modules of ``groups of type $E_7$'', a set of (semi)simple Lie groups sharing common structural/geometrical properties as encapsulated by the FTS axioms. 

\begin{definition}[Freudenthal triple system \cite{Brown:1969}] An FTS is axiomatically defined  as a finite dimensional
vector space $\mathfrak{F}$ over a field $\mathds{F}$ (not of characteristic
2 or 3), such that:

\begin{enumerate}
\item $\mathfrak{F}$ possesses a non-degenerate antisymmetric bilinear form $%
\{x, y\}.$

\item $\mathfrak{F}$ possesses a symmetric four-linear form $q(x,y,z,w)$
which is not identically zero.

\item If the ternary product $T(x,y,z)$ is defined on $\mathfrak{F}$ by $%
\{T(x,y,z), w\}=q(x, y, z, w)$, then
\begin{equation}  \label{eq:def}
3\{T(x, x, y), T(y,y,y)\}=\{x, y\}q(x, y, y, y).
\end{equation}
\end{enumerate}
\end{definition}

Groups of type $E_7$ are defined in terms of the ``automorphisms'' of the triple product.  

\begin{definition}[Automorphism group \cite{Brown:1969}]
The \emph{automorphism} group of an FTS is defined as the subset of invertible $%
\mathds{F}$-linear transformations  preserving the quartic and quadratic
forms:
\begin{equation}
\textrm{\emph{Aut}}(\mathfrak{F}):=\{\sigma \in \textrm{\emph{Iso}}_{\mathds{F}}(\mathfrak{F})|\{\sigma
x,\sigma y\}=\{x,y\},\;q(\sigma x)=q(x)\}.  \label{eq:brownfts}
\end{equation}%
\end{definition}

Note, the conditions $\{\sigma x,\sigma y\}=\{x,y\}$ and $q(\sigma x)=q(x)$
immediately imply $\sigma$ acts as an automorphism of the triple product,
\begin{equation}\label{eq:Taut}
T(\sigma x, \sigma y, \sigma z)=\sigma T(x, y, z),
\end{equation}
hence the name.

The conventional concept of matrix rank may be generalised to Freudenthal triple systems in a natural and $\Aut(\mathfrak{F})$-invariant manner. 
\begin{definition}[The FTS Rank \cite{Ferrar:1972, Krutelevich:2004}] 
The rank of an arbitrary element $x\in\mathfrak{F}$ is  defined by:
\be\label{eq:FTSrank}
\begin{array}{lll}
\textrm{\emph{Rank}} (x)=0&\Leftrightarrow\left\{\begin{array}{l} x=0\end{array}\right.\\[7pt]
\textrm{\emph{Rank}} (x)=1&\Leftrightarrow\left\{\begin{array}{l} x\not=0\\\Upsilon_x(y)=0\;\forall y
\end{array}\right. \\[15pt]
\textrm{\emph{Rank}} (x)=2&\Leftrightarrow\left\{\begin{array}{l} \exists y\;\textrm{s.t.}\;\Upsilon_x(y)\not=0,\\
T(x,x,x)=0\end{array}\right.\\[15pt]
\textrm{\emph{Rank}} (x)=3&\Leftrightarrow\left\{\begin{array}{l} T(x,x,x)\not=0\\q(x)=0\end{array}\right.\\[15pt]
\textrm{\emph{Rank}} (x)=4&\Leftrightarrow\left\{\begin{array}{l} q(x)\not=0\end{array}\right.\\
\end{array}
\ee
where we have defined $\Upsilon_x(y):=3T(x,x,y)+x\{x,y\}x$.
\end{definition}

The ranks partition $\FTS$ and are manifestly invariant under $\Aut(\mathfrak{F})$. Note,  they are self-consistent and ordered in the sense that,
\be
\bes
x=0 &\Rightarrow\Upsilon_x(y)=0;\\
\Upsilon_x(y)=0 &\Rightarrow T(x,x,x)=0;\\
T(x,x, x)=0 &\Rightarrow q(x)=0.\\
\end{array}
\ee

The rank condition can be understood in terms of the representation theory of $\Aut(\FTS)$. Recall, $\FTS$ constitutes an $\Aut(\FTS)$-module. Define,
\be
\bes
\Upsilon: \FTS\times\FTS &\rightarrow \Hom_\F(\FTS, \FTS)\\
(x,y)&\mapsto \Upsilon_{x,y}
\end{array}
\ee
where
\be
 \Upsilon_{x,y}(z):=3T(x,y,z)+\frac{1}{2}\{x,z\}y+\frac{1}{2}\{y,z\}x.
\ee
Then $ \Upsilon$ belongs to $\mathfrak{Aut}(\FTS)$, the Lie algebra of $\Aut(\FTS)$. That is, $\Upsilon$ is the projection onto the adjoint in $\Sym^2(\FTS)$. This follows from the observation   that  $\mathfrak{Aut}(\mathfrak{F})$ is
given by all $\phi \in \Hom_{\mathds{F}}(\mathfrak{F}, \mathfrak{F}
)$ such that $q(\phi x,x,x,x)=0$ and $\{\phi x,y\}+\{x,\phi y\}=0$  for all $x,y\in \mathfrak{F}$,
as is easily verified \cite{Borsten:2011nq}.

\begin{lemma}
The $\mathds{F}$-linear map $\Upsilon_x:\mathfrak{F}\rightarrow \mathfrak{F}$
defined by
\begin{equation}  \label{eq:adj}
\Upsilon_x(y)=3T(x,x,y)+\{x,y\}x
\end{equation}
is in $\mathfrak{Aut}(\mathfrak{F})$. Linearizing \eref{eq:adj} with
respect to $x$ implies that $\Upsilon_{x,y}:\mathfrak{F}\rightarrow
\mathfrak{F}$ defined by
\begin{equation}
\Upsilon_{x,y}(z)=6T(x,y,z)+\{x,z\}y+\{y, z\}x
\end{equation}
is in $\mathfrak{Aut}(\mathfrak{F})$. \end{lemma}
Note, $\Upsilon_{x,y}$ is a manifestly $\Aut(\mathfrak{F})$-covariant expression for the Freudenthal product $x\wedge y$  given in \cite{Yokota:2009}.
To establish this simple result, note
\be
\{\Upsilon_z(x), y\}+\{x, \Upsilon_z(y)\}=0
\ee
 follows directly from the antisymmetry and symmetry of $\{x, y\}$ and  $q(x,y,z,w)=\{T(x,y,z), w\}$, respectively. The second condition, $q(\phi x,x,x,x)=0, \forall \phi \in \Aut(\mathfrak{F})$, is  satisfied since
\be
q(x,x,x, \Upsilon_z(x))=3\{T(x,x,x),T(z,z,x)\}+\{z, x\}\{T(x,x,x), z\}
\ee
vanishes due to the defining FTS relation \eref{eq:def}.

Similarly, $T$ is the projection onto $\FTS$ in $\Sym^3(\FTS)$, as confirmed by \eref{eq:Taut}, while $q$ is by definition the singlet in $\Sym^4(\FTS)$.

\subsection{The three-qubit FTS}\label{sec:3qubitfts}

Consider the  three-qubit pure states,
\be
\ket{\psi}=a_{ABC}\ket{ABC}, \quad A,B,C=0,1
\ee
in $\mathcal{H}_{ABC}=\C^2\otimes\C^2\otimes\C^2$.
For notational clarity, we will use both $e^{ABC}$ and $\ket{ABC}$ interchangeably to denote  the computational basis vectors.

\begin{definition}[Three-qubit FTS] The FTS of three qubits is defined by,
\be
\{e^{ABC}, e^{A'B'C'}\}:=\varepsilon^{AA'}\varepsilon^{BB'}\varepsilon^{CC'}
\ee
and
\be
\begin{array}{l}
q(e^{A_1B_1C_1}, e^{A_2B_2C_2}, e^{A_3B_3C_3}, e^{A_4B_4C_4}):=\\[5pt]
\displaystyle\frac{1}{4!}\sum_{\textrm{\emph{perms}}\{1,2,3,4\}}\varepsilon ^{A_1A_2}\varepsilon ^{A_3A_4}\varepsilon ^{B_1B_2}\varepsilon^{B_3B_4}\varepsilon ^{C_1C_4}\varepsilon ^{C_2C_3}\\[20pt]
=\frac{1}{6}\left\{\varepsilon ^{A_1A_2}\varepsilon ^{A_3A_4}\varepsilon ^{B_1B_2}\varepsilon^{B_3B_4}\varepsilon ^{C_1C_4}\varepsilon ^{C_2C_3}\right.\\[4pt]
~~   +\,\varepsilon ^{A_1A_2}\varepsilon ^{A_4A_3}\varepsilon ^{B_1B_2}\varepsilon ^{B_4B_3}\varepsilon ^{C_1C_3}\varepsilon^{C_2C_4}\\[4pt]
~~   +\,\varepsilon ^{A_1A_3}\varepsilon ^{A_2A_4}\varepsilon ^{B_1B_3}\varepsilon^{B_2B_4}\varepsilon ^{C_1C_4}\varepsilon ^{C_3C_2}\\[4pt]
 ~~   +\,\varepsilon ^{A_1A_3}\varepsilon^{A_4A_2}\varepsilon ^{B_1B_3}\varepsilon ^{B_4B_2}\varepsilon ^{C_1C_2}\varepsilon^{C_3C_4}\\[4pt]
 ~~   +\,\varepsilon ^{A_1A_4}\varepsilon ^{A_2A_3}\varepsilon ^{B_1B_4}\varepsilon^{B_2B_3}\varepsilon ^{C_1C_3}\varepsilon ^{C_4C_2}\\[4pt]
 \left.~~ +\varepsilon ^{A_1A_4}\varepsilon^{A_3A_2}\varepsilon ^{B_1B_4}\varepsilon ^{B_3B_2}\varepsilon ^{C_1C_2}\varepsilon^{C_4C_3}\right\}.
\end{array}
\ee
\end{definition}
Here $\varepsilon^{AA'}$ is the $\SL_A(2, \mathds{C})$-invariant antisymetric $2 \times 2 $ tensor, where $\varepsilon^{01}=1$. With these definitions $\mathcal{H}_{ABC}$ forms an FTS, as can be verified by checking  \eref{eq:def}.  In fact, this FTS is based on an underlying Jordan algebra $\J_{ABC}\cong\C\oplus\C\oplus\C$. For a  detailed discussion of this construction the reader is referred to \cite{Borsten:2009yb}. 

The automorphism group is 
\be
\SL(2, \mathds{C})\times \SL(2, \mathds{C})\times\SL(2, \mathds{C})\rtimes S_3,
\ee
where $S_3$ denotes the  three-qubit permutation group. The automorphism invariant  rank conditions of \eref{eq:FTSrank} are given explicitly by the following tensors.

For a state
$
\ket{\psi}=a_{ABC}\ket{ABC}, 
$
 the quartic norm $q(\psi)$ is given  by
\begin{equation}\label{eq:q}
\begin{array}{lll}
q(\psi)&=2\det\gamma^{A}=2\det\gamma^{B}=2\det\gamma^{C}\\
       &=-2\Det a_{ABC},
\end{array}
\end{equation}
where $\Det a_{ABC}$ is Cayley's hyperdeterminant \cite{Cayley:1845, Miyake:2002} and we have introduced the three symmetric matrices $\gamma^A,\gamma^B$, and $\gamma^C$ defined by,
\begin{equation}
\begin{array}{lll}\label{eq:ABCgammas}
(\gamma^{A})_{A_{1}A_{2}}&=\varepsilon^{B_{1}B_{2}}\varepsilon^{C_{1}C_{2}}a_{A_{1}B_{1}C_{1}}a_{A_{2}B_{2}C_{2}}, \\
(\gamma^{B})_{B_{1}B_{2}}&=\varepsilon^{C_{1}C_{2}}\varepsilon^{A_{1}A_{2}}a_{A_{1}B_{1}C_{1}}a_{A_{2}B_{2}C_{2}}, \\
(\gamma^{C})_{C_{1}C_{2}}&=\varepsilon^{A_{1}A_{2}}\varepsilon^{B_{1}B_{2}}a_{A_{1}B_{1}C_{1}}a_{A_{2}B_{2}C_{2}},
\end{array}
\end{equation}
transforming respectively as a $\rep{(3,1,1), (1,3,1), (1,1,3)}$  under $\SL(2,\mathds{C}) \times \SL(2,\mathds{C}) \times \SL(2,\mathds{C})$. Explicitly,
\begin{equation}\label{eq:gammasex}
\begin{array}{lll}
\gamma^{A}&=
\left(\begin{array}{cc}
2(a_{0}a_{3}-a_{1}a_{2}) &  a_{0}a_{7}-a_{1}a_{6}+a_{4}a_{3}-a_{5}a_{2}\\
a_{0}a_{7}-a_{1}a_{6}+a_{4}a_{3}-a_{5}a_{2}  & 2(a_{4}a_{7}-a_{5}a_{6})
\end{array}\right), \\[16pt]
\gamma^{B}&=
\left(\begin{array}{cc}
2(a_{0}a_{5}-a_{4}a_{1}) & a_{0}a_{7}-a_{4}a_{3}+a_{2}a_{5}-a_{6}a_{1}\\
a_{0}a_{7}-a_{4}a_{3}+a_{2}a_{5}-a_{6}a_{1} & 2(a_{2}a_{7}-a_{6}a_{3})
\end{array}\right), \\[16pt]
\gamma^{C}&=
\left(\begin{array}{cc}
2(a_{0}a_{6}-a_{2}a_{4}) &  a_{0}a_{7}-a_{2}a_{5}+a_{1}a_{6}-a_{3}a_{4}\\
a_{0}a_{7}-a_{2}a_{5}+a_{1}a_{6}-a_{3}a_{4} & 2(a_{1}a_{7}-a_{3}a_{5})
\end{array}\right),
\end{array}
\end{equation}
where we have made the decimal-binary conversion 0, 1, 2, 3, 4, 5, 6, 7 for 000, 001, 010, 011, 100, 101, 110, 111.
In the same notation the Hyperdeterminant is,
\begin{equation}\label{eq:CayleyHyperdeterminant}
\begin{array}{lll}
\Det a  
=&\phantom{-\ }a_{0}^2 a_{7}^2 + a_{1}^2 a_{6}^2+  a_{2}^2 a_{5}^2 + a_{3}^2 a_{4}^2 \\
&-2\,(a_{0}a_{1}a_{6}a_{7} +a_{0}a_{2} a_{5}a_{7} +a_{0}a_{4}a_{3}a_{7}\\
&\phantom{-2\,(\ }+a_{1}a_{2}a_{5}a_{6} +a_{1}a_{3}  a_{4}a_{6} +a_{2}a_{3}a_{4}a_{5}) \\
&+4\,(a_{0}a_{3}a_{5}a_{6}+ a_{1}a_{2}a_{4}a_{7}).
\end{array}
\end{equation}
The triple product is given by
\be
\ket{T(\psi)}=T(a)_{ABC}\ket{ABC}, 
\ee
where $T(a)_{ABC}$ may be written in three equivalent ways
\begin{equation}\label{eq:Tofgamma}
\begin{array}{lll}
T_{A_3B_1C_1}=\varepsilon^{A_1A_2}a_{A_1B_1C_1}(\gamma^{A})_{A_{2}A_{3}},\\
T_{A_1B_3C_1}=\varepsilon^{B_1B_2}a_{A_1B_1C_1}(\gamma^{B})_{B_{2}B_{3}},\\
T_{A_1B_1C_3}=\varepsilon^{C_1C_2}a_{A_1B_1C_1}(\gamma^{C})_{C_{2}C_{3}},
\end{array}
\end{equation}
each of which makes the identity 
\be
q(\psi)=\{{T(\psi)}, {\psi}\}
\ee
manifest. 
Finally, $\Upsilon_{\psi}(\phi)$ for an arbitrary state $\ket{\phi}=b_{ABC}\ket{ABC}$  is given by
\be
\ket{\Upsilon_{\psi}(\phi)}=\Upsilon_{ABC}\ket{ABC}
\ee
where
\be
\begin{array}{lll}\label{eq:up}
\Upsilon_{ABC}=&-\varepsilon^{A_1A_2}b_{A_2BC}(\gamma^A)_{AA_1}\\
&-\varepsilon^{B_1B_2}b_{AB_2C}(\gamma^B)_{BB_1}\\
&-\varepsilon^{C_1C_2}b_{ABC_2}(\gamma^C)_{CC_1}.
\end{array}
\ee

\subsection{SLOCC entanglement classification}

 The concept of SLOCC  equivalence was introduced in \cite{Bennett:1999, Dur:2000}. Two states  lie in the same SLOCC-equivalence class if and only if they may be transformed into one another with some \emph{non-zero probability} using LOCC operations. For more on LOCC operations and entanglement the reader is refereed to \cite{Vedral:1997, Plenio:2007, Horodecki:2007} and the references therein.
 The crucial observation is that since LOCC cannot create entanglement any two SLOCC-equivalent states must necessarily possess the same entanglement, irrespective of the particular measure used. It is this property which make the SLOCC paradigm so suited to the task of classifying entanglement. 
 
Restricting our attention to pure states, two $n$-qubit states
are SLOCC-equivalent if and only if they are related by an element of
$\SL_1(2, \mathds{C})\times \SL_2(2, \mathds{C})\times \ldots\SL_n(2, \mathds{C})$ \cite{Dur:2000}, which will be referred to as the \emph{SLOCC-equivalence group}. The Hilbert space is partitioned into equivalence classes or orbits under the SLOCC-equivalence group. Hence, for the $n$-qubit system the space of SLOCC-equivalence classes  is given by,
\begin{equation}\label{eq:SLOCCeqiv}
\frac{{\mathds C}^2\otimes{\mathds C}^2\ldots\otimes\C^2}{\SL_1(2, \mathds{C})\times \SL_2(2, \mathds{C})\times \ldots\SL_n(2, \mathds{C})}.
\end{equation}
This  is the space of physically distinct entanglement classes; the SLOCC entanglement classification amounts to understanding \eref{eq:SLOCCeqiv}.

In the case of three qubits the SLOCC-equivalence group coincides with the three-qubit FTS automorphism group and the space of entanglement classes  \eref{eq:SLOCCeqiv} is determined by the ranks as in  \autoref{tab:merge} \cite{Borsten:2009yb}.  All states of a given rank 1, 2 or 3  are SLOCC-equivalent while the set of rank 4 states constitute a $\dim_\C=1$ family of equivalent states parametrised by $q(\Psi)$. 
More specifically, the entanglement classes and their (unnormalised) representative states are as follows:  
\begin{enumerate}
\item (Rank 1) Totally separable states $A$-$B$-$C$, 
\be
\begin{array}{c}
\ket{000}
\end{array}
\ee

\item (Rank 2) Biseparable states $A$-$BC$, $B$-$CA$, $C$-$AB$,
\be
\begin{array}{c}
\ket{000}+\ket{011}\\
\ket{000}+\ket{101}\\
\ket{000}+\ket{110}\\
\end{array}
\ee

\item  (Rank 3) Totally entangled  $W$ states, 
\be
\begin{array}{c}
\ket{011}+\ket{101}+\ket{110}
\end{array}
\ee

\item (Rank 4) One-parameter family of totally entangled GHZ states 
\be
\begin{array}{c}
a\ket{000}-\ket{011}-\ket{101}-\ket{110}
\end{array}
\ee 
where $q(\psi)=8a$.
\end{enumerate}
Since the rank conditions are ordered by implication, so are the entanglement classes. The rank 4 GHZ class is regarded as maximally entangled in the sense that it has non-vanishing quartic norm. Note, the three rank 2 classes  collapse in to a single class  since the three matrices  $\gamma^{A, B, C}$, given in \eref{eq:ABCgammas},  are rotated into each other under  the three-qubit permutation group.

\begin{table}
\begin{tabular}{cccM{c}M{c}cc}
\hline
 \multirow{2}{*}{Class} 	& \multirow{2}{*}{Rank} 	& \multirow{2}{*}{Representative state}& \multicolumn{2}{c}{\textrm{FTS rank condition}}                \\
\cline{4-5}
                        			&                       			& & \textrm{vanishing}                     & \textrm{non-vanishing}  \\
\hline
 Null                   		& 0  		             		& $-$& \Psi  				                  & -                     \\
 $A$-$B$-$C$            	& 1  		             		& $\ket{000}$& 3T(\Psi,\Psi,\Phi)+\{\Psi,\Phi\}\Psi & \Psi 		           \\
 $A$-$BC$               	& 2a 		            		& $\ket{000}+\ket{011}$& T(\Psi,\Psi,\Psi) 					  & \gamma^A               \\
 $B$-$CA$               	& 2b 		             		& $\ket{000}+\ket{101}$& T(\Psi,\Psi,\Psi) 					  & \gamma^B               \\
 $C$-$AB$               	& 2c 		             		& $\ket{000}+\ket{110}$& T(\Psi,\Psi,\Psi)                    & \gamma^C	           \\
 $W$                      		& 3  		             		& $\ket{011}+\ket{101}+\ket{110}$& q(\Psi)							  & T(\Psi,\Psi,\Psi) 	  \\
 GHZ                    		& 4  		            		& $a\ket{000}-\ket{011}-\ket{101}-\ket{110}$& -                                    & q(\Psi)              \\
\hline
\end{tabular}
\caption{Three-qubit entanglement classification  as according to the FTS rank system.\label{tab:merge}}
\end{table}

\section{Non-local games}\label{sec:nonlocal}
A non-local game, as introduced in \cite{1313847}, consists of \emph{players} (Alice, Bob, Charlie\ldots), who act cooperatively in order to win, and  a \emph{referee} who coordinates the game. The players may collectively decide on a strategy before the game commences. Once it has begun they may no longer communicate. Whether or not the players win is determined by the referee. To begin the referee randomly selects one question, from a known fixed set $\mathcal{Q}$, to be sent to each player. The players know only their own questions. Each player must then send back a response from the set of answers, denoted $\mathcal{A}$. The referee determines whether the players win using the set of sent questions and received answers according to some predetermined rules. These rules are known to the players before the game gets under way so that they may attempt to devise a winning strategy.   

For the  three-player game \cite{Watrous:2006} the questions sent to Alice, Bob and Charlie, denoted respectively by $r,s$ and $t$,   are taken from the set $\mathcal{Q}=\{0, 1\}$. However, the referee ensures that $rst\in\{000, 110, 101, 011\}$ with a uniform distribution and the players are aware of this. The answers $a, b, c$,  sent back by Alice, Bob and Charlie,   are  elements of $\mathcal{A}=\{0, 1\}$.  The players win if $r\vee s\vee t = a\oplus b\oplus c$, where $\vee$ and $\oplus$ respectively denote disjunction and addition mod 2, i.e for question sets $rst=000, 011, 101$ and $110$ the answer set $abc$ must satisfy $a\oplus b\oplus c = 0,1,1$ and $1$, respectively.

In the quantum version, Alice, Bob and Charlie each possess a qubit, which they may manipulate locally. Any entanglement  shared by the three qubits  can potentially be used to the players advantage. However, before examining how this works  let us consider first how well the players can do classically,  unassisted by entanglement.

What is the best possible classical {deterministic\footnote{We need only consider this case here since, for non-local games, the best  winning probability possible using a  deterministic strategy  is  an upper bound on the best winning probability possible using a probabilistic strategy \cite{1313847, Watrous:2006}.} strategy}? A deterministic strategy amounts to specifying three functions, one for each player,  from the question set $\mathcal{Q}$ to the answer set $\mathcal{A}$, 
\be
\begin{array}{lll}
a: \mathcal{Q}\rightarrow \mathcal{A};& \quad r\mapsto a(r),\\
b: \mathcal{Q}\rightarrow \mathcal{A};& \quad s\mapsto b(s),\\
c: \mathcal{Q}\rightarrow \mathcal{A};& \quad t\mapsto c(t),
\end{array}
\ee
 The condition that the players win may then be written as,
\be
\begin{array}{lll}
a(0)\oplus b(0)\oplus c(0) &= 0,\\
a(1)\oplus b(1)\oplus c(0) &= 1,\\
a(1)\oplus b(0)\oplus c(1) &= 1,\\
a(0)\oplus b(1)\oplus c(1) &= 1.
\end{array}
\ee
This implies that the best one can do is win $75\%$ of the time; the four equations cannot be simultaneously satisfied as can be seen by adding them mod 2 \cite{Watrous:2006}.  On the other hand, the simple strategy that ``everyone always answers $1$'' satisfies three of the four equations so that the $75\%$ upper bound is actually met.

Can this be bettered when equipped with an entangled resource? The answer is a resounding yes: by sharing a GHZ state,
\be\label{eq:winstate}
\ket{\Psi}=\frac{1}{{2}}\left(\ket{000}-\ket{011}-\ket{101}-\ket{110}\right),
\ee
 they can always win \cite{Watrous:2006}.  

The winning quantum strategy is remarkably simple.   If a player receives the question ``0''  they measure their qubit in the computational basis $\{\ket{0}, \ket{1}\}$. If a player receives the question ``1''  they measure their qubit in the Hadamard basis $\{\left(\ket{0}+\ket{1}\right)/\sqrt{2}, \left(\ket{0}-\ket{1}\right)/\sqrt{2}\}$.  The measurement outcome is sent back as their answer. By symmetry we need only consider the two cases  $rst=000$ and $rst = 011$. 
(1)  $rst=000$: All players measure in the computational basis. From \eref{eq:winstate} it is clear that only an odd number of 0's can appear $\Rightarrow$ $a\oplus b \oplus c=0$. Always win.
(2) $rst=011$: Alice measures in the computational basis, while Bob and Charlie measure in the Hadamard  basis. Consulting the locally rotated state,
\be
\mathds{1}\otimes H\otimes H \ket{\Psi}=\frac{1}{2}\left(\ket{001}+\ket{010}-\ket{100}+\ket{111}\right),
\ee
where  $H$ is the Hadamard matrix, it is clear that only an even number of 0's can appear
$\Rightarrow$ $a\oplus b \oplus c=1$. Always win.
Hence, using the GHZ entangled resource \eref{eq:winstate} Alice, Bob and Charlie can win certainty, outdoing the best possible classical strategy. 

\section{Freudenthal ranks: GHZ vs. $W$}

We will now show that the FTS rank conditions imply that there is no local strategy utilising  a rank 3  state that wins with certainty. Similarly, the optimal rank 2 state strategy   falls short of the rank 3 case implying that the winning probabilities are ordered by  rank.

A local strategy corresponds to choosing six unitary rotations, $R_r, S_s, T_t$, where $r,s,t=0,1$, one pair each for Alice, Bob and Charlie. Let
\be
\ket{\psi^{rst}}= \psi_{ABC}^{rst}\ket{ABC}= R_r \otimes S_s \otimes T_t\ket{\psi},
\ee
where $\ket{\psi}$ is the initial shared state. Note, for notational convenience will shall use the decimal expression for both $rst$ and $ABC$, so that, for example, the amplitudes of $\ket{\psi^{000}}$ are $\psi^{0}_{0}, \psi^{0}_{1},\ldots, \psi^{0}_{7}$.

Since it is assumed $\ket{\psi}$ is a rank 3 state we have  $T(\psi)\not=0, \Det \psi =0$, which implies  
\be\label{eq:Wrank1}
T(\psi^{rst})\not=0, \Det \psi^{rst} =0.
\ee

Let us now assume that there is in fact a strategy that wins with certainty. For $rts=0$ this implies 
\be
\psi^{0}_{7,1,2,4}=0.
\ee
Hence
\be
\Det \psi^0 = 4 \psi^{0}_{0} \psi^{0}_{3} \psi^{0}_{5}\psi^{0}_{6}
\ee
and
\numparts
\begin{eqnarray}\label{eq:0gamma1}
\gamma^{A}(\psi^0)&=&
\left(\begin{array}{cc}
2\psi^{0}_{0}\psi^{0}_{3} & 0\\
0 & -2\psi^{0}_{5}\psi^{0}_{6}
\end{array}\right)
, \\\label{eq:0gamma2}
\gamma^{B}(\psi^0)&=&
\left(\begin{array}{cc}

2\psi^{0}_{0}\psi^{0}_{5} & \\
 & -2\psi^{0}_{6}\psi^{0}_{3}
\end{array}\right)
, \\\label{eq:0gamma3}
\gamma^{C}(\psi^0)&=&
\left(\begin{array}{cc}
2\psi^{0}_{0}\psi^{0}_{6}&0  \\
0 & -2\psi^{0}_{3}\psi^{0}_{5}
\end{array}\right)
,
\end{eqnarray}
\endnumparts
which together imply that one and only one of $\psi^{0}_{0,3,5,6}$ must be vanishing in order that the rank condition $T(\psi^0)\not=0, \Det \psi^0 =0$ be satisfied.

Similarly, for $rst=i=3,5,6$ we have 
\be
\psi^{i}_{0,3,5,6}=0,
\ee
\be
\Det \psi^i = 4 \psi^{i}_{7} \psi^{i}_{1} \psi^{i}_{2}\psi^{i}_{4}
\ee
and
\numparts
\begin{eqnarray}\label{eq:igamma1}
\gamma^{A}(\psi^i)&=&
\left(\begin{array}{cc}
-2\psi^{i}_{2}\psi^{i}_{1}  & 0\\
0 & 2\psi^{i}_{7}\psi^{i}_{4}
\end{array}\right)
, \\\label{eq:igamma2}
\gamma^{B}(\psi^i)&=&
\left(\begin{array}{cc}
-2\psi^{i}_{1}\psi^{i}_{4}  &0 \\
0 & 2\psi^{i}_{7}\psi^{i}_{2}
\end{array}\right)
, \\\label{eq:igamma3}
\gamma^{C}(\psi^i)&=&
\left(\begin{array}{cc}
-2\psi^{i}_{4}\psi^{i}_{2}  &0  \\
0 & 2\psi^{i}_{7}\psi^{i}_{1}
\end{array}\right)
,
\end{eqnarray}
\endnumparts
which, again, imply that one and only one of $\psi^{i}_{7,1,2,4}$ must be vanishing.

Note, by the covariance of the rank condition we have
\numparts
\begin{eqnarray}
\gamma^{A}(\psi^{rst})&=&e^{i(\phi_s+\lambda_t)}R_r\gamma^{A}(\psi)R_{r}^{T}, \\
\gamma^{B}(\psi^{rst})&=&e^{i(\theta_r+\lambda_t)}S_s\gamma^{B}(\psi)S_{s}^{T}, \\
\gamma^{C}(\psi^{rst})&=&e^{i(\theta_r+\phi_s)}T_t\gamma^{C}(\psi)T_{t}^{T},
\end{eqnarray}
\endnumparts
where $\det R_r =e^{i\theta_r}$, $\det S_s =e^{i\phi_s}$, $\det T_t =e^{i\lambda_t}$, so that
\numparts
\begin{eqnarray}
e^{-i(\phi_s+\lambda_t)}\gamma^{A}(\psi^{rst})&=&e^{-i(\phi_{s'}+\lambda_{t'})}\gamma^{A}(\psi^{rs't'}), \\
e^{-i(\theta_r+\lambda_t)}\gamma^{B}(\psi^{rst})&=&e^{-i(\theta_{r'}+\lambda_{t'})}\gamma^{B}(\psi^{r'st'}), \\
e^{-i(\theta_{r}+\phi_{s})}\gamma^{C}(\psi^{rst})&=&e^{-i(\theta_{r'}+\phi_{s'})}\gamma^{C}(\psi^{r's't}).
\end{eqnarray}
\endnumparts
Hence, from equations \eref{eq:0gamma1} through \eref{eq:0gamma3} and \eref{eq:igamma1} through \eref{eq:igamma3}  we obtain the following set of 12 conditions:
\numparts
\begin{eqnarray}
|\psi_{0}^{0}||\psi_{3}^{0}|&=|\psi_{2}^{3}||\psi_{1}^{3}|, \\
|\psi_{5}^{0}||\psi_{6}^{0}|&=|\psi_{7}^{3}||\psi_{4}^{3}|, \\\nonumber
\\
|\psi_{2}^{5}||\psi_{1}^{5}|&=|\psi_{2}^{6}||\psi_{1}^{6}|, \\
|\psi_{7}^{5}||\psi_{4}^{5}|&=|\psi_{7}^{6}||\psi_{4}^{6}|, \\\nonumber
\\
|\psi_{0}^{0}||\psi_{5}^{0}|&=|\psi_{1}^{5}||\psi_{4}^{5}|, \\
|\psi_{5}^{0}||\psi_{6}^{0}|&=|\psi_{7}^{5}||\psi_{2}^{5}|, \\\nonumber
\\
|\psi_{1}^{3}||\psi_{4}^{3}|&=|\psi_{1}^{6}||\psi_{4}^{6}|, \\
|\psi_{7}^{3}||\psi_{2}^{3}|&=|\psi_{7}^{6}||\psi_{2}^{6}|,\\\nonumber
\\
|\psi_{0}^{0}||\psi_{6}^{0}|&=|\psi_{4}^{6}||\psi_{2}^{6}|, \\
|\psi_{3}^{0}||\psi_{5}^{0}|&=|\psi_{7}^{6}||\psi_{1}^{6}|, \\\nonumber
\\
|\psi_{4}^{3}||\psi_{2}^{3}|&=|\psi_{4}^{5}||\psi_{2}^{5}|, \\
|\psi_{7}^{3}||\psi_{1}^{3}|&=|\psi_{7}^{5}||\psi_{1}^{5}|. 
\end{eqnarray}
\endnumparts
 Under the rank condition that one and only one of each of $\psi^{0}_{0,3,5,6}$ and $\psi^{i}_{7,1,2,4}$ must be vanishing this set of equations has no solution, yielding a contradiction. Hence, $p(\textrm{rank } 3)<p(\textrm{rank } 4)=1$ as claimed.
 
Using a similar  logic one can show that $p(\textrm{rank } 2)<p(\textrm{rank } 3)$.  While the details offer no particular insight beyond the previous case,  the argument does rely on two simple, but perhaps not immediately obvious, observations. First, adopting the  optimal  GHZ strategy for a $W$ state the players win with probability 7/8, as a quick calculation will confirm.  Second,   the rank 2 conditions imply that for any rank 2 state one and only one of $\gamma^{A, B, C}$ is non-vanishing. This follows from the identity,\begin{equation}
(\gamma^A)_{A_{1}A_{2}}(\gamma^C)_{C_{1}C_{2}}=\varepsilon^{B_2B_1}a_{A_1B_1C_1}T_{A_2B_2C_2}+\varepsilon^{B_1B_2}a_{A_2B_2C_1}T_{A_1B_1C_2},
\end{equation}
which implies that if $T_{ABC}=0$ then there is at most one non-vanishing $\gamma$, while the non-vanishing of $\Upsilon$ implies at least one non-zero $\gamma$ as can be seen from \eref{eq:up}.

Using these conditions it can be shown directly that $p(\textrm{rank } 2)<p(\textrm{rank } 3)$. However, a more illuminating way to proceed follows  from the fact that one and only one non-vanishing $\gamma$  implies that the state is biseparable.  Let us consider with out loss of generality (by the symmetry of the game) the $A$-$BC$ split with representative state $\ket{\psi}_A\ket{\phi}_{BC}$. Defining the observables $A_r=R^{\dagger}_{r} \sigma_z R_{r}$, $A_r=S^{\dagger}_{s} \sigma_z S_{s}$ and $C_t=T^{\dagger}_{t}\sigma_z T_{t}$, we note that the expectation value of
\be
E:=A_0B_0C_0-A_0B_1C_1-A_1B_0C_1-A_1B_1C_0
\ee
is four times the difference between the probability of winning and losing the game. In the biseparable case we can therefore use a Tsirelson type argument \cite{cirel1980quantum}. Let 
\be
\begin{array}{ll}
S&=\bra{\psi}{}\bra{\phi}E\ket{\psi}\ket{\phi}\\
&=\bra{\phi} B_0 (a_0 C_0-a_1C_1)- B_1 (a_1 C_0+a_0C_1)\ket{\phi}\\
\end{array}
\ee
where $a_r = \bra{\psi}A_r\ket{\psi}\in[-1,1]$. Then
\be
\begin{array}{ll}
S&\leq ||[B_0 (a_0 C_0-a_1C_1)- B_1 (a_1 C_0+a_0C_1)]\ket{\phi}||\\[4pt]
&\leq ||B_0 (a_1C_1-a_0 C_0)\ket{\phi}||+|| B_1 (a_1 C_0+a_0C_1)\ket{\phi}||\\[4pt]
&\leq ||\mathds{1} \otimes(a_1 C_1-a_0 C_0)\ket{\phi}||+||  \mathds{1} \otimes (a_1  C_0+a_0 C_1)\ket{\phi}||\\[4pt]
&= ||a_1 \ket{\phi_1}-a_0 \ket{\phi_0}||+|| a_1  \ket{\phi_0}+a_0 \ket{\phi_1}||,
\end{array}
\ee
where $\ket{\phi_t}=\mathds{1}\otimes C_t \ket{\phi}$. Since $||\phi_t||\leq1$ and $a_r \in[-1,1]$ we have
\be
S\leq\sqrt{2-2a_0a_1\Re \braket{\phi_0}{\phi_1}}
+\sqrt{2+2a_0a_1\Re \braket{\phi_0}{\phi_1}}
\ee
which is just the usual Tsirelson bound $S\leq 2\sqrt{2}$. Hence, 
\be
p(\textrm{rank } 2) \leq \frac{1}{2}+\frac{1}{2\sqrt{2}} < p(\textrm{rank } 3).
\ee

Finally, as the above analysis suggests, played with  a rank 2 biseparable state,  $\ket{000}+\ket{011}$ say,  where Alice decides to always answer ``0'',  the three-player game is equivalent (bit-flipping the rules) to the Clauser, Horne, Shimony and Holt (CHSH) two-qubit game  \cite{Clauser:1969,1313847}. Hence, there is indeed a local strategy that wins with probability \[\frac{1}{2}+\frac{1}{2\sqrt{2}}\] and it is  trivially true that $p(\textrm{rank } 1)=3/4<p(\textrm{rank } 2)$.

\section{Further work}

We have shown that the optimal success rates  when sharing a three-qubit  resource are strictly ordered according as the rank of the state used: 
\be
3/4=p(\textrm{rank } 1)<p(\textrm{rank } 2)<p(\textrm{rank } 3)<p(\textrm{rank } 4)=1.
\ee
We conclude that the inherent ordering of the Freudenthal rank conditions is reflected physically by the increasing advantage acquired with respect to the Freudenthal rank  of the  entangled state used. It would be interesting to understand to what extent this observation applies beyond the three-qubit case. Indeed, by reverting back to the Jordan algebraic perspective it is possible to generalise the basic features of the FTS to $n$ qubits \cite{Borsten:2013qqq}\footnote{Essentially this is simply a particular way of organising the conventional SLOCC entanglement classification using SLOCC-equivalence group covariants. However, while much is know about the four qubit case \cite{Verstraete:2002, Briand:2003a,Chterental:2007,Borsten:2010db,PhysRevA.83.052330,PhysRevA.86.042332,PhysRevA.87.062305}, the complete covariant SLOCC classification is only known up to three qubits.}. The two-qubit case is rather simple: there are two ranks corresponding to a single orbit of separable sates and a one-parameter family of entangled states. The CHSH game \cite{1313847} somewhat trivially reflects  the ordering of the ranks. In the four-qubit case, on the other hand, the complete set of ranks\footnote{By ``complete set'' we mean the minimal number of independent covariants required to characterise the entanglement classification.} is not even known and, moreover, one would anticipate them to be only partially ordered, since there are four independent SLOCC-equivalence group invariants \cite{Briand:2003a}. Nonetheless, it would be interesting, given a complete set of ranks, to attempt to identify a minimal set of non-local games that would ``experimentally verify'' this expected partial order. This non-trivial task is very much left as an open problem. 

Returning briefly now to the three-qubit case in hand, we remark that the non-local properties   of the $W$  and GHZ states may also be compared  using the  sheaf-theoretic framework of \cite{abramsky2011sheaf, 2012arXiv1203.1352A}.  In this case one applies in both instances the  winning GHZ strategy described in \autoref{sec:nonlocal}. The resulting GHZ model is shown to be \emph{strongly contextual}, admitting no global section, while the $W$ model is merely \emph{contextual} \cite{abramsky2011sheaf, 2012arXiv1203.1352A}. It would be interesting to understand to what extent this sheaf-theoretic take on non-locality, and its associated   notions of strong contextuality etc,  can be understood in terms of FTS ranks and  more generally the conventional SLOCC perspective on entanglement classes.

Finally, we note in passing that the Freudenthal ranks  determine the degree of supersymmetry preserved by the single-centre extremal black hole solutions of $\mathcal{N}=8$ supergravity \cite{Ferrara:1997uz, Borsten:2008wd, Borsten:2010aa}, suggesting an admittedly tenuous link between non-local games and Killing spinor equations. 

\ack
LB would like to thank MJ Duff and J Henson for helpful discussions. The work of LB is supported by an Imperial College Junior Research Fellowship.

\end{document}